\documentclass[%
 aip,
rsi,%
 amsmath,amssymb,
 reprint,%
]{revtex4-1}

\usepackage{graphicx}
\usepackage{dcolumn}
\usepackage{bm}
\usepackage{upgreek}

\begin{document}

\preprint{AIP/123-QED}

\title[]{Dynamically tunable metamaterial analogue of electromagnetically induced transparency with graphene in the terahertz regime}

\author{Tingting Liu}\email{ttliu@hust.edu.cn}
\affiliation{School of Electronic Information and Communications, Huazhong University of Science and Technology, Wuhan 430074, People's Republic of China}


\begin{abstract}
A novel mechanism to realize dynamically tunable electromagnetically induced transparency (EIT) analogue in the terahertz (THz) regime is proposed. By putting the electrically controllable monolayer graphene under the dark resonator, the amplitude of the EIT resonance in the metal-based metamaterial can be modulated substantially via altering the Fermi level of graphene. The amplitude modulation can be attributed to the change in the damping rate of the dark mode caused by the recombination effect of the conductive graphene. This work provides an alternative way to achieve tunable slow light effect and has potential applications in THz wireless communications.
\end{abstract}

\maketitle
The electromagnetically induced transparency (EIT) effect is of great interest in many important applications, such as slow light, switching, and nonlinear devices. This effect was first demonstrated in three-level atomic systems, where the destructive interference between two radiative transitions creates a sharp transparency window within a broad absorption spectrum.\cite{harris1990Nonlinear} However, quantum EIT requires complicated experimental handlings, such as stable gas lasers and low temperature environment, which constraints its practical implementation in chip-scale applications. Recently, metamaterial analogues of EIT based on the near field coupling between the bright and dark resonators\cite{zhang2008Plasmon,liu2009plasmonic,liu2012electromagnetically}, has been intensively investigated because of the flexible design and easy implementation. These EIT metamaterials bring the original quantum effect into the realm of classical optics and show great potentials in developing novel plasmonic devices.

In practice, the ability to dynamically tune EIT effect has attracted enormous attention. For most EIT analogues based on metallic materials with a fixed spectral response, it is difficult to manipulate the EIT response without changing the geometries or modifying the supporting substrates. For this, the EIT metamaterials are integrated with active materials, such as semiconductors\cite{Gu2012active,xu2016frequency}, superconductors\cite{cao2013plasmon} and nonlinear materials\cite{zharov2003nonlinear}. Alternatively, graphene can be an excellent candidate for designing tunable devices in the terahertz (THz) regime. Its surface conductivity can be tuned by shifting the Fermi level, which may be potentially varied from -1 to 1 eV by chemical doping or electrostatic gating\cite{liu2011chemical,hu2015broadly}. Moreover, the graphene-based metamaterial devices show high modulation speed, compared with other counterparts.\cite{liu2011graphene,li2014ultrafast} Hence recent studies have proposed a variety of graphene-based metamaterial to generate controllable EIT analogues.\cite{cheng2013dynamically,ding2014tuneable,xia2016dynamically,he2016terahertz,zhao2016graphene,yao2016dynamically} However, the surface conductivity of the discrete graphene resonator is difficult to be tuned, restricting the practical applications.

In this letter, we propose a hybrid THz EIT metamaterial by integrating the controllable monolayer graphene into the metal-based metamaterial composed of a cut wire (CW) and a pair of split-ring resonators (SRRs). A significant amplitude modulation of the EIT resonance is demonstrated by shifting the Fermi level of graphene. Based on the investigation of the coupled harmonic oscillator model and electrical field distributions, the modulation can be attributed to the change in the damping rate of the dark mode caused by the recombination effect of the conductive graphene . The proposed EIT metamaterial achieves the dynamically tunable group delay and show promising applications in developing compact slow light devices.

\begin{figure*}[htbp]
\centering
\includegraphics[scale=0.9]{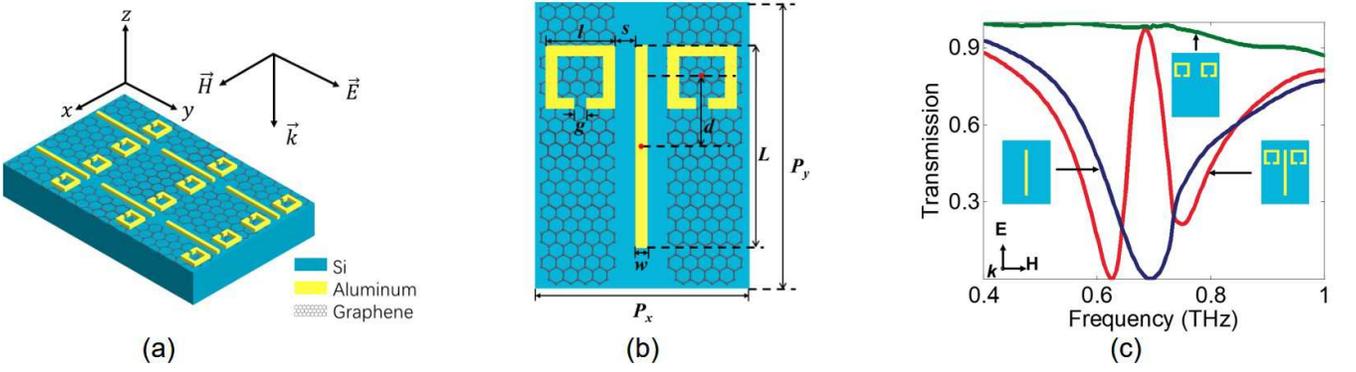}
\caption{\label{Fig:1} (a) schematic of the proposed EIT metamaterial and the incident light polarization configuration; (b) the unit cell of the proposed structure. The geometric parameters are $P_x=80$ $\upmu$m, $P_y=120$ $\upmu$m, $L=85$ $\upmu$m, $l=29$ $\upmu$m, $s=7$ $\upmu$m, $g=5$ $\upmu$m, $w=5$ $\upmu$m and $d=28$ $\upmu$m, respectively. (c) Simulated transmission spectra for the CW array, the SRR-pair array and the proposed EIT array excited by a y-polarized incident electric field.}
\end{figure*}

The schematic geometry of our proposed EIT metamaterial is depicted in FIG. 1(a). The unit cell is composed of a pair of SRRs symmetrically placed on the left and right sides of a CW, and the monolayer graphene is deposited on the bottom of the SRR-pair. The specific geometrical parameters are described in FIG. 1(b). The SRR-pair and CW are both made of 200 nm-thick aluminum (Al). The THz characteristics of Al are modeled by the complex permittivity $\varepsilon_{Al}=\varepsilon_{\infty}-\omega_p^2/(\omega^2+i\omega\gamma)$, where $\omega_p=2.24\times10^{16}$ rad/s is the plasma frequency and $\gamma=1.22\times10^{14}$ rad/s is the damping constant.\cite{ordal1985optical} The silicon (Si) substrate is assumed to be semi-infinite and lossless with refractive index $n_{Si}=3.42$. In the THz regime, the complex valued surface conductivity of graphene can be modeled by the Drude-like expression, $\sigma=ie^2E_F/(\pi\hbar^2(\omega+i\tau^{-1}))$, where $e$ is the charge of an electron, $E_F$ is the Fermi level in graphene, $\hbar$ is the reduced Planck's constant and $\tau$ is the relaxation time.\cite{xiao2016tunable,xiao2016spectrally,xiao2017strong} The relaxation time $\tau=\mu E_F/(ev_F^2)$ is dependent on the Fermi level $E_F$, carrier mobility $\mu$ and Fermi velocity $v_F$.\cite{zhang2014coherent,zhang2015towards} According to the experimental measurements, we set the Fermi velocity as $1.1\times10^6$ m/s, and the mobility as 3000 cm$^2/$V$\cdot$s in this work.\cite{zhangyb2005experimental,jnawali2013observation} By shifting the Fermi level $E_F$, the surface conductivity of graphene can be dynamically modulated. The numerical simulations are further conducted using finite-difference time-domain method.

The simulated transmission spectra of the three metamaterial structures, including a CW array, a SRR-pair array and the proposed EIT array are presented in FIG. 1 (c). The incident electric field $E$ is oriented along the y axis. The CW array exhibits a localized surface plasmon (LSP) resonance at 0.68 THz, whereas the LC resonance in SRR-pair at the same frequency is inactive due to the structural symmetry. By arranging them within a unit cell in the proposed structure, the two types of resonators are coupled with each other and a typical EIT spectral response with a significant transparency peak is observed within a broad absorption background. The CW is excited directly by y-polarized incidence serving as the bright mode, and LC resonance in SRR-pair is excited by the near field coupling effect serving as the dark mode. It is the destructive interference between the bright and dark modes that results in the EIT response with a sharp transparency window at 0.68 THz.

To realize the dynamic tunability of the EIT metamaterial, we integrate the monolayer graphene into the unit cell and investigate the effect of the Fermi level on the transmission properties, as shown in FIG. 2 (a). When there is no graphene, a significant transparency peak is obtained at 0.68 THz with a transmission amplitude of 97$\%$. As the Fermi level of graphene gradually increases from 0.1 eV to 0.4 eV, the transmission amplitude of EIT peak decreases from 67$\%$ to 22$\%$ without notable frequency shift. With further increasing Fermi level to the maximum value of 0.8 eV, the transparency peak vanishes and only a resonance dip remains at 0.68 THz in the transmission spectrum. Finally, the transmission spectrum of the EIT metamaterial loses the transparency window and turns into a typical LSP resonance. The tunability of the EIT metamaterial highlights the realization of active control of EIT response in THz devices without reconstructing the structure geometries.

\begin{figure}[htbp]
\centering
\includegraphics[scale=0.45]{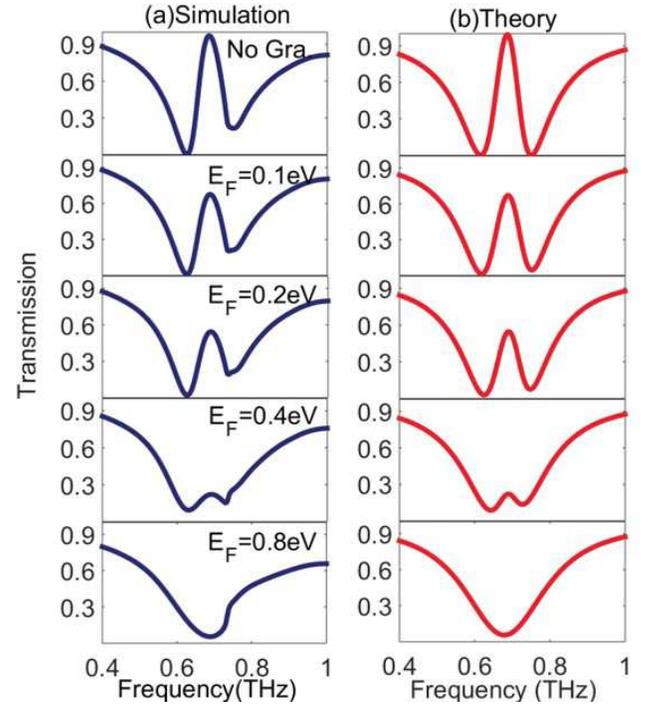}
\caption{\label{Fig:2}(a) Simulated transmission spectra and (b) corresponding theoretical fitting results for the proposed EIT metamaterial with various values of Fermi level of graphene.}
\end{figure}

To clarify the underlying physical mechanism of the tunable EIT effect, we adopt a classical coupled harmonic oscillator model to analyze the coupling effect between the bright and dark modes in the EIT metamaterial. The susceptibility of the EIT metamaterial $\chi$ can be derived as \cite{liu2009plasmonic}
\begin{equation}
    \chi(\omega)=\chi_r+i\chi_i\propto\frac{\omega-\omega_1+i\frac{\gamma_1}{2}}{(\omega-\omega_0+i\frac{\gamma_0}{2})
    (\omega-\omega_1+i\frac{\gamma_1)}{2}-\frac{\kappa^2}{4}},
\label{eq1}
\end{equation}
where $\gamma_1$ and $\gamma_2$ are the damping rates of the bright and dark modes; $\omega_0=2\pi\times0.68$ THz and $\omega_1=\omega_0+\delta$ are resonance frequencies of the bright and dark modes; $\kappa$ is the coupling coefficient between the two modes. Since the absorption is proportional to the imaginary part of susceptibility $\chi_i$ in the system, the relationship between the transmission and $\chi_i$ can be obtained through\cite{ding2014tuneable}
\begin{equation}
  T(\omega)=1-g\chi_i(\omega),
\label{eq1}
\end{equation}
where $g$ is a geometric parameter describing the coupling strength between the bright mode and the incident electric field. Using the two equations, the analytical fittings can be conducted for the simulated transmission spectra under various Fermi levels. As shown in FIG. 2(b), the analytical results exhibit a good agreement with the simulations. The corresponding fitting parameters with various Fermi levels, including $\gamma_1$, $\gamma_2$, $\kappa$ and $\delta$ are depicted in FIG. 3. During the tuning process, $\gamma_1$, $\kappa$ and $\delta$ do not change notably, whereas  $\gamma_2$ increases significantly from 0.0005 THz for the case without graphene to 0.28 THz for the case with Fermi level of 0.8 eV. Hence, it can be inferred that the tunability of the proposed EIT metamaterial mainly results from the variation of $\gamma_2$, the damping rate of the dark mode. In the proposed EIT metamaterial, the monolayer graphene placed under the SRR-pair, bridges the two ends of each split as a conductive layer. With the increase of Fermi level of graphene, the increasing conductivity of graphene gradually enhances the losses in SRR-pair and weaken the destructive interference between the LC resonance in SRR-pair and LSP resonance in CW. Finally, the damping rate of the dark mode is large enough to suppress the dark mode excitation and constrain the appearance of the EIT resonance.

\begin{figure}[htbp]
\centering
\includegraphics[scale=0.45]{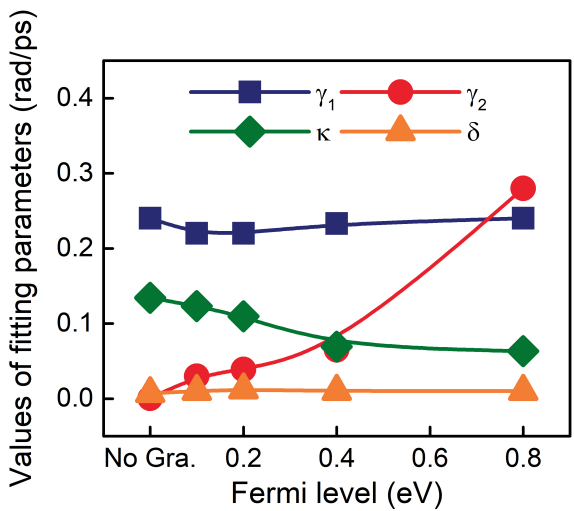}
\caption{\label{Fig:3}The variation of fitting parameters $\gamma_1$, $\gamma_2$, $\kappa$ and $\delta$ of the coupled harmonic oscillator model as a function of the Fermi level of graphene.}
\end{figure}

To gain a deeper insight into the modulation mechanism of the proposed EIT analogue, the distributions of the simulated normally polarized electric field at the EIT resonance frequency are investigated. FIG. 4 provides the x-y plane electric field ($E_z$) distributions at 0.68 THz with different Fermi levels of graphene. For the case without graphene, the LC dark mode in SRR-pair is indirectly excited by the coupling effect with the LSP bright mode in CW, and the electric field in CW is suppressed by the destructive interference. When the graphene is integrated, the recombination effect of the conductive graphene leads to the redistribution of electrical field in the EIT metamaterial. When the Fermi level is 0.2 eV, the excitation of the LC dark mode in SRR-pair is partly suppressed due to the increased graphene conductivity and the electric field is localized on both SRR-pair and CW. When the Fermi level increases to 0.8 eV, the large conductivity of graphene enhances the recombination effect, suppresses the excitation of the dark mode, and eliminates the electric field in SRR-pair. Hence the destructive interference between the bright and dark modes almost disappears. It can be concluded that the dynamically tunability of the EIT-like effect results from the manipulation of graphene conductivity via altering its Fermi level.
\begin{figure*}[htbp]
\centering
\includegraphics[scale=0.7]{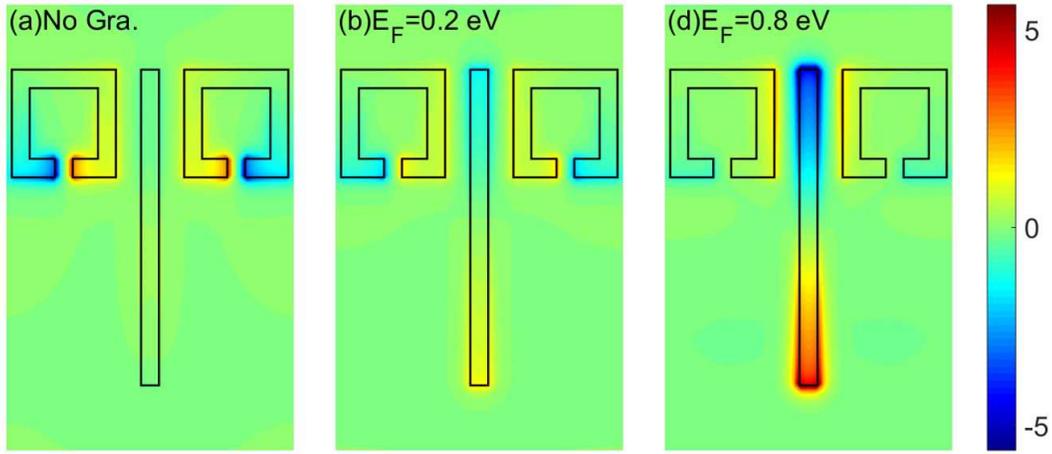}
\caption{\label{Fig:4}Distributions of normalized x-y plane electric field ($E_z$) of the proposed EIT structure for the cases of (a) without graphene (b) $E_F=0.2$ eV (c) $E_F=0.8$ eV.}
\end{figure*}

The EIT effect has the ability to slow down the speed of light. The dynamically tunable transmission spectra in FIG. 2 imply that the proposed EIT metamaterial can realize the active control of slow light effect. FIG. 5 depicts the dependency of the phase shift ($\psi$) and the group delay ($\tau_g$) on the Fermi level of graphene. By tuning the Fermi level of graphene, the normal phase dispersion, as well as group delay within the transparency window experiences the modulation within a large range, which reveals great potential in the active control of slow light applications. When there is no graphene, the EIT metamaterial shows the largest phase shift at the EIT peak and reaches the maximum group delay $\tau_g=4.6$ ps. As the Fermi level of graphene increases, both the phase change and group delay gradually decrease, following the same tendency with that of the transparency window of transmission spectra shown in FIG. 2. Under the different Fermi levels as 0.1, 0.2 and 0.4 eV, the group delays $\tau_g$ at the EIT peak are 3.2, 2.6, and 0.5 ps, respectively. As the $E_F$ increases to 0.8 eV, the slow light characteristic of EIT metamaterial disappears. Therefore, the proposed EIT metamaterial can achieve tunable slow light effect, which shows great promising in developing compact THz slow light devices.
\begin{figure}[htbp]
\centering
\includegraphics[scale=0.45]{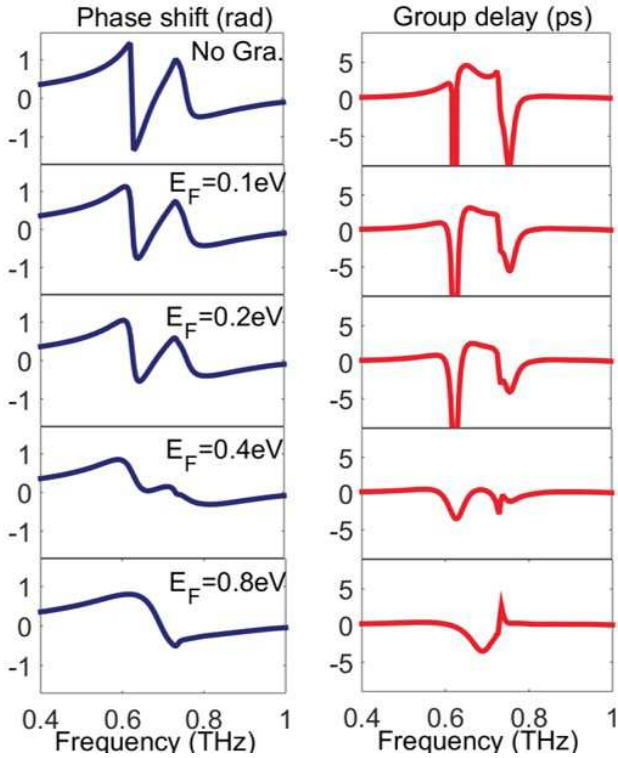}
\caption{\label{Fig:5}The phase change ($\psi$)(left column) and the corresponding group delay ($\tau_g$) (right column) for the proposed EIT structure with various values of Fermi level of graphene.}
\end{figure}

In summary, we investigate a dynamically tunable EIT metamaterial design in THz regime by integrating the monolayer graphene into the unit cell consisting of a CW and a SRR-pair. By altering the Fermi level of the integrated graphene, the significant modulation of transmission amplitudes can be achieved. To explain the modulation mechanism, the EIT resonance is described based on the classical coupled harmonic oscillator model, and the tunable characteristic is attributed to the increasing damping rate of the LC dark mode arising from increasing graphene conductivity via altering its Fermi level. The electric field distributions are further investigated to validate our interpretation. In addition, we also investigate the tunable group delays of the EIT metamaterial, which may provide meaningful guidance for compact, tunable slow light devices. Our proposed EIT metamaterial can pave the way for innovative design and realization of dynamically controllable THz slow light devices, which may find potential applications in future THz wireless communications.

%

\end{document}